\documentclass{aa}
\usepackage{psfig}
\usepackage{graphics}

\begin{document}

   \title{Detection of a radio halo in the Virgo cluster\thanks{Based
on observations with the Effelsberg 100-m telescope operated by the
Max-Planck-Institut f\"ur Radioastronomie (MPIfR), Bonn, Germany} }

   \author{B. Vollmer\inst{1,2}, W. Reich\inst{2},
          \and
          R. Wielebinski\inst{2}}

   \offprints{B. Vollmer: bvollmer@astro.u-strasbg.fr}

   \institute{
     CDS, Observatoire astronomique de Strasbourg, UMR 7550, 11, rue de l'universit\'e, 67000 Strasbourg, France \and
             Max-Planck-Institut f\"ur Radioastronomie, Auf dem H\"ugel 69,
             53121 Bonn, Germany
             }

   \date{Received ; accepted }

\abstract{New Effelsberg 1.4~GHz observations of the central $10\degr \times 10\degr$ of the Virgo cluster
are presented. NVSS data are used to subtract point sources from our map.
During the data reduction process special care is taken (i) to disentangle emission from the
North Polar Spur from emission from the Virgo cluster, (ii) to disentangle emission from the strong
M87 sidelobes
from emission from the Virgo cluster, and (iii) to correct for non-linear ground emission due to
the long scans. We detect a low surface brightness radio halo with a flux density of
$5\pm1.5$~Jy centered close to the
elliptical galaxy M86. This halo is much weaker than that observed in the Coma cluster.
It is reminiscent of a past interaction between the intracluster medium of M86 and
a low density gas, belonging most probably to the Virgo cluster.
}

\titlerunning{The Virgo cluster radio halo}
   \maketitle
%

\section{Introduction \label{sec:intro}}

The location of the largest diffuse radio emission sources in the local Universe are galaxy clusters.
A very hot (several $10^{7}$~K) and tenuous ($n \sim 10^{-4}$~cm$^{-3}$) gas pervades the whole cluster.
This gas can be directly observed in X-rays (see e.g. Ebeling et al. 1996). The existence of a cluster-wide
diffuse radio emission indicates that there are additional, important components of the
intracluster medium (ICM): large-scale
magnetic fields and relativistic electrons. In general these extended, diffuse radio sources
are divided into two subgroups (see e.g. Giovannini \& Feretti 2002a):
(i) cluster halos have low surface brightness, are centered on the cluster center, and show a regular shape.
Their prototype is Coma~C (Deiss et al. 1997). They have a steep spectrum and show a negligible degree
of polarization. Their typical size is $\sim 1$~Mpc. (ii) Relic sources also show low surface brightness
and steep spectra, but in contrast to radio halos they are elongated structures that are found at the periphery
of the cluster. In addition, their degree of polarization is high. A3667 (R\"{o}ttgering et al. 1997) represents
an extreme member of this group.

Diffuse, extended radio sources are rare. Giovannini et al. (1999), using a complete cluster sample,
found that 5\% of clusters have radio halos and 6\% show relic sources. The X-ray luminosity of clusters hosting a
diffuse radio source is significantly higher than that of clusters without such a source (Giovannini \& Feretti 2002b).
The number of known diffuse radio sources has significantly increased to about 40 in the last years.
Typical radio powers of diffuse sources are $10^{24} - 10^{25}$~W\,Hz$^{-1}$ at 1.4~GHz. The strength of the magnetic
fields calculated using equipartition between the relativistic electrons and the magnetic field is
$0.1 - 1$~$\mu$G.

There is strong evidence that diffuse radio sources are related to cluster mergers. Significant substructure
in the X-ray emission is detected in clusters hosting a radio halo/relic (B\"{o}hringer \& Sch\"{u}cker 2002).
At present a merger scenario could not be clearly excluded for a cluster showing a radio halo (Giovannini \& Feretti 2002a).
The formation of halos and relics is tightly linked to the dynamical behaviour of the cluster.
Whereas the magnetic fields and relativistic electron populations are built during the cluster formation and subsequent
evolution, the energy supply for the maintenance of the diffuse radio sources is due to recent gravitational
interactions between the cluster and an infalling galaxy group or cluster.

One of the best studied cluster radio sources is Coma (see e.g. Deiss et al. 1997), which is located at
$\sim$100~Mpc distance. Its extent on the sky is about $1\degr$. In order to detect a large-scale, diffuse emission
in a galaxy cluster with a single-dish telescope, additional high angular resolution
observations at the same wavelength must be available to identify point
sources, which have to be subtracted. It must also be taken into account that the region
of interest should not be too large, because the low surface brightness of diffuse radio sources implies
huge amount of observation time at high frequencies.
In addition, observations of large areas may have problems in the determination of a constant baselevel.
Because of these reasons, it is only now that we attempt to detect a diffuse radio source in the
Virgo cluster with the Effelsberg 100-m telescope. The Virgo cluster, with a distance of
17~Mpc\footnote{We use a Hubble constant of $H_{0}=70$~km\,s$^{-1}$\,Mpc$^{-1}$.}, is
the nearest galaxy cluster on the northern hemisphere. We cover the central $10\degr \times 10\degr$.

Virgo is a dynamically young cluster, i.e. the galaxy distribution is not symmetric (see e.g. Schindler et al. 1999).
Several subgroups of galaxies are located closely in projected distance and velocity to the Virgo cluster.
The galaxy distribution shows mainly a north-south and an east-west elongation. The ICM distribution observed
with ROSAT in X-rays (B\"{o}hringer et al. 1994) shows a symmetric component around M87 and an asymmetric,
low surface brightness emission elongated in north-south direction.
The main part of this asymmetric component comes from the galaxy group around the elliptical galaxy M49 located
in the south of M87. A small X-ray halo is also associated with M86, a big elliptical galaxy located to the west
close to M87.

These asymmetries imply a high dynamical activity. At the same time the Virgo cluster hosts a central cooling flow
(White \& Sarazin 1988). Since it is thought that a major cluster merger disrupts a central cooling flow
(see White et al. (1993) for the Coma cluster), the existence
of a symmetric X-ray halo and a cooling flow around M87 thus implies that the Virgo cluster did not experience
a major merger in the recent past. Nevertheless, this might happen in the near future (Tully \& Shaya 1984).

From the central cD galaxy M87 a distorted radio jet emanates, surrounded by a radio minihalo (Andernach et al. 1979,
Rottman et al. 1996, Owen et al. 2000).
Its linear extent is about 80~kpc. This minihalo is directly related to the radio jet of M87 and can thus
not be qualified as a radio halo as defined above.

\begin{figure}
\resizebox{\hsize}{!}{\includegraphics{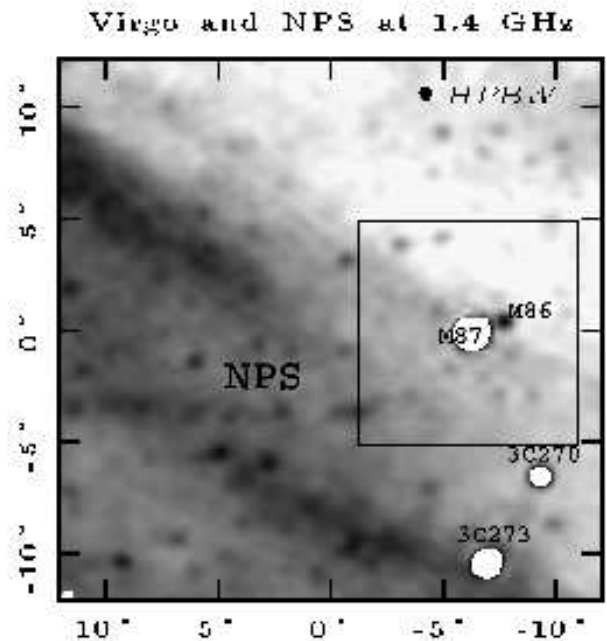}}
 \caption{Section of the 1.4~GHz Stockert continuum survey showing a part of the North Polar Spur close to the Virgo cluster
with Virgo~A or M87 at its center.
The greyscale range is between 3.3~K (white) and 3.7~K (black). Intensities above 3.7~K are clipped (shown white).
The map is in the equatorial coordinate system and centered at $\rm 12^{h}57^{m}18.^{s}4, 12\degr24\arcmin24\arcsec$ (J2000)
 or $6\degr$ to the east of Virgo~A. The $10\degr \times 10\degr$ field observed with the Effelsberg telescope is indicated.}
\label{fig:NPS}
\end{figure}

We have attempted to detect diffuse radio emission on a Mpc scale in the Virgo cluster at 1.4~GHz with the
Effelsberg 100-m telescope.
When conducting a search for a presumably weak radio halo from the Virgo cluster,
the presence of the quite bright radio source Virgo~A (3C274, M87) must be taken into account.
At 1.4~GHz the peak flux density is about 150~Jy when measured with the
Effelsberg telescope. Such a strong signal produces antenna sidelobes
within the area of the cluster what must be adequately taken into account. Another
complication arises from the North Polar Spur (NPS), an intense Galactic
radio shell, which is likely the old remnant from a nearby supernova explosion. Its minimum distance was previously estimated to be about 100~pc.
Its diameter is about $116\degr \pm 4\degr$ (Berkhuijsen et al. \cite{elly}).
The NPS is seen in projection  towards the south-eastern area of the Virgo cluster. Virgo~A is close to
the outer periphery of the NPS as shown in Fig.~\ref{fig:NPS}, where a section of the 1.42~GHz northern--sky
Stockert survey (Reich \& Reich \cite{pr}) is displayed to illustrate the unfavourable situation.
However, this absolutely calibrated Stockert survey data may be helpful in a separation
of the different contributions.

The observations, data reduction, and the results are described in Sect.~\ref{sec:data} followed by the
discussion in Sect.~\ref{sec:discussion} and the our conclusions (Sect.~\ref{sec:conclusions}).


\section{Observations, data reduction and results \label{sec:data}}

We used the Effelsberg 100-m telescope at 1.4~GHz to map a $10\degr \times 10\degr$ field
centered on Virgo~A in the equatorial coordinate system. The observations were started in
January 2001 and completed in February 2002. Only data from night-time observations were used to
avoid any influence by solar sidelobe emission. The field was mapped in orthogonal directions
in the equatorial coordinate system by
moving the telescope with a speed of $\rm 4\degr$/min. The subscan separation was
$4\arcmin$, which provides full sampling for a beam width (HPBW) of $9\farcm4$. The large size of
the field required to split the observations into several subfields. In total about
two full coverages and in addition some incomplete sections were obtained in both scanning
directions. The integration time per pixel was between 4 and 5~sec.

Effelsberg continuum and polarization observations with the L-band prime-focus receiver, their
reduction and calibration were already described by Uyan{\i}ker et al. (\cite{bu98}). In brief,
the receiver has two channels for the amplification of the left and right circularly polarized
components and is tunable within a frequency range from 1.29~GHz to 1.72~GHz. Cooled HEMT amplifiers
reduce the system noise to about 30~K. The drift stability of the system is extremely high that
it is operated in total-power mode without any switching. Weather conditions have in practice no influence
on the quality of the data. Only in extreme cases of snow and rain fall the noise increases and the
baselines are affected. Observations are interrupted in these rare cases. However, impulsive or low level
broadband interference was sometimes noted, which needs careful editing.
For the observations presented here two IF-polarimeters were used in parallel with center frequencies
at 1.395~GHz and 1.408~GHz. The bandwidth of each polarimeter was 14~MHz. The cross-talk between
the left and right circular components
was found to differ for both subbands and each of them needs to be separately corrected, following
in principle the method described by Uyan{\i}ker et al. (\cite{bu98}), where the cross-talk of the
total-power channels into the polarization channels was canceled. The amount of cross-talk is a few
percent of the total intensity signal and is a tribute to the
wide frequency range where the receiver can be used.

The calibration of the data relies on the nearby strong radio source 3C286 as the primary
calibrator, assuming 14.4~Jy and a percentage polarization of 9.3\% at $33\degr$ polarization angle.
Daily calibration
parameters were calculated from the average of all observed calibration sources during a night.
The conversion factor from the Jy/beam-scale into main beam brightness temperature was determined
by Uyan{\i}ker et al. (\cite{bu98}) to 2.12$\pm$0.02~K/(mJy/beam).
We did not make a spectral correction for the frequency difference of both subbands used.

The raw data were acquisited at a rate of 16~msec in $2\times4$~channels. Continuous calibration was
applied, where an additional calibration signal was injected for 50\% of time.  The calibration signal
was also used to correct for any phase drift in the IF-polarimeter. The data were interpolated  onto
a 4$\arcmin$ grid along the scanning direction, the same as for the scan separation. Linear baselines were subtracted,
which were calculated from a few datapoints at both ends of each scan. This is a standard procedure
for single--dish observations, however, only a first step in our case because of the significant influence
of the NPS in about half of the map. In addition a systematic negative curvature remains, which reflects
the non--linear ground radiation picked up by the telescope's far--sidelobes. This effect may mask or at least reduce any
existing symmetric halo centered on Virgo~A we aim to detect. For Stokes U and Q linear baseline settings to zero at the edges
of the observed map are a compromise as well, since polarized Galactic emission exists everywhere, but
requires absolute measurements to get its large-scale level. As an alternative, which is mostly used in synthesis
telescope observations, the average of a map in U or Q is defined as the zero level.

The observed scans were arranged into maps, edited by standard procedures for spiky interference and
baseline distortions of individual scans using an unsharp masking procedure (Sofue \& Reich \cite{sofue}).
Correction for cross--talk of the polarization channels and conversion
of the raw data into the main beam brightness scale finalizes the raw data reduction. All total intensity maps
as well as all U and Q maps were combined using the PLAIT procedure (Emerson \& Gr\"ave \cite{emerson88}),
which includes a destriping algorithm by weighting down Fourier components along the individual scanning
direction of each map. The maps from both subbands were checked for differences, which might show residual
interference effects or in the case of U and Q indicate significant Faraday rotation of the polarized emission.
However, the differences in U and Q were within the noise and we averaged the maps from both channels in the
same way as the total intensity maps.

From the final maps in U and Q we measured a rms-noise of $\sim 3$~mK~$\rm T_{b}$ in flat regions.
Measuring the rms-noise from the total intensity map is difficult as the map is clearly limited by confusion,
which is about 7~mJy/beam or 15~mK $\rm T_{b}$ at 1.4~GHz for the Effelsberg telescope (Uyan{\i}ker et
al. \cite{bu99}) at a resolution of $9\farcm4$. The measured noise is in the confusion range.

\begin{figure}
\resizebox{\hsize}{!}{\includegraphics{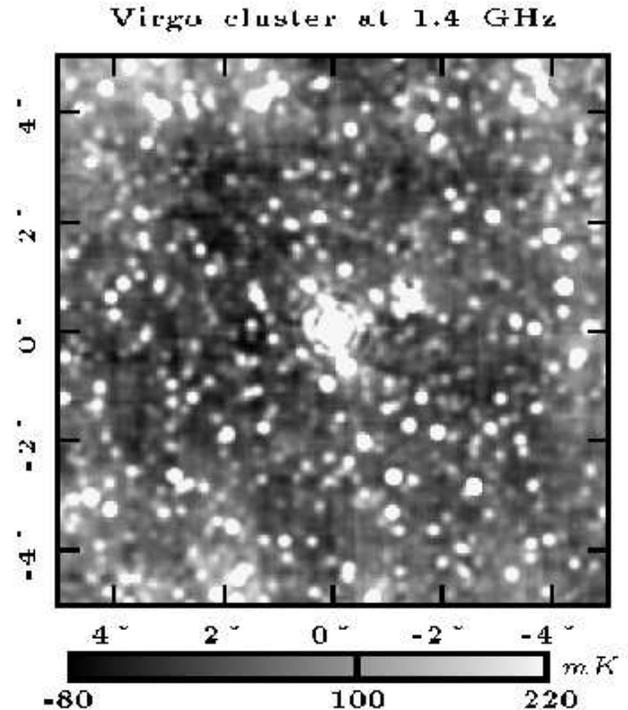}}
 \caption{Effelsberg total intensity map of the Virgo cluster field centered on Virgo~A (M87) at  $\rm 12^{h}
30^{m}49.^{s}4, +12\degr 23'28.0''$(J2000). The resolution is $9\farcm4$.}
   \label{fig:imap}
\end{figure}

\begin{figure}
\resizebox{\hsize}{!}{\includegraphics{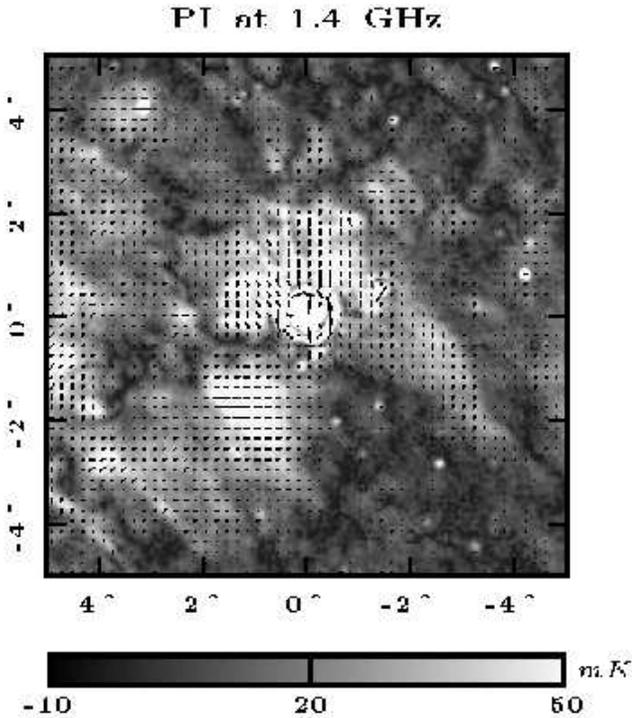}}
 \caption{Effelsberg polarized intensity map of the Virgo cluster field centered on Virgo~A (M87) at $\rm 12^{h}
30^{m}49.^{s}4, +12\degr 23'28.0''$(J2000). The resolution is $9\farcm4$. The lines show the E vectors.}
   \label{fig:pimap}
\end{figure}

We show the observed I map in Fig.~\ref{fig:imap}.
Some systematic total intensity decrease towards the map center is due to residual nonlinear ground radiation.
The total intensity map shows clear radial sidelobe structures centered on Virgo~A due to the four
subreflector support legs of the telescope. These limit the dynamic range of the final map to about $-28$dB
close to Virgo~A after averaging all maps which were observed
at different parallactic angles. We tried to further lower these sidelobe responses by
taking into account the contribution from the individual maps to the central region. Five maps contribute to the central
region which have mean parallactic angles in the range between $37\degr$ and $-30\degr$. Differences of the
individual maps compared to the mean map were used to blank regions with strong sidelobes.
The fairly uniform ring-like, but weaker diffraction sidelobes remained almost unchanged.
In an iterative process the dynamic range could be improved. This method was previously applied
to remove the sidelobe response from the subreflector support legs from CAS~A, which is stronger than M87,
in the Effelsberg 1.4~GHz Galactic plane survey by Reich et al. (\cite{pr97}, their Figs.~1 and 2).
The issue how residual sidelobe structures might affect the search for a radio halo
is further discussed in Sect.~\ref{sec:sidelobes}.

Polarized intensities and polarization angles were calculated from the U and Q maps after they were offset-adjusted
so that their mean is zero. The final map of polarized intensity is shown in Fig.~\ref{fig:pimap}.
The main beam response of U and Q from Virgo~A must be
considered as residual instrumental polarization and was clipped. In the ring of first sidelobes some
residual instrumental polarization is visible which
is on a level of less than 0.1\% of the maximum total intensity from Virgo~A. Weak radial structures centered
on Virgo~A have an instrumental origin. Away from Virgo~A patchy
coherent polarization structures are visible with maximum polarized intensities of about 70~mK (south-east of Virgo~A).
A number of filamentary structures are running mostly in the direction of the NPS filaments. The NPS is
known to be a highly polarized feature in the sky (Spoelstra 1972a, 1972b). The Effelsberg map misses some fraction
of this large-scale emission as already mentioned in the context of baseline setting.
Since the observed polarized emission is clearly of Galactic
origin, we do not consider these data any more in the following analysis.

\subsection{Source subtraction}

In order to increase the sensitivity for diffuse emission, which is clearly
limited by confusion of unresolved compact sources, we make use of the NVSS (Condon et al.
\cite{con}), a sensitive source survey at 1.4~GHz carried out with the VLA. The survey maps
show sources down to about 1~mJy at $\sim45\arcsec$ angular resolution and are insensitive
to all structures exceeding about 4$\farcm$8. We combined all NVSS data for the observed $10\degr \times 10\degr$
field and fitted compact sources down to about 1~mJy with an elliptical Gaussian. Residual sources
where tried to be fitted with a more robust, circular Gaussian in a second run.
No attempt was made to fit and subtract M87. We constructed a
map from the fitted sources which is free of the low-level instrumental artifacts visible in
the original NVSS survey maps. This map was convolved to the same angular resolution as the
Effelsberg observations. We found the scaling factor for the source map by fitting the slope
of an intensity-versus-intensity plot for a number of subsections of the map. The scatter
was found to be less than 10\%. The convolved and scaled source map was then subtracted from the
Effelsberg map. A few residual sources left from the source fitting procedure were removed
afterwards. In some cases residuals at the location of strong sources remain. The reasons are
a complex source structure not well fitted by a Gaussian, variability, pointing differences between
the Effelsberg and the VLA maps or scaling differences in case of strong sources. This procedure
reduces the noise-level of the total intensity map to about 4~mK $\rm T_{b}$ ($3.1$~mJy/beam)
when convolved to a resolution of $12\arcmin$. In the following all
maps are convolved to a resolution of $12\arcmin$.

\subsection{Total intensity baselevel setting}

As already mentioned, a systematic negative curvature towards the center of the total intensity
map remains, which reflects
the non--linear ground radiation picked up by the telescope's far--sidelobes, masking or at least reducing
any existing symmetric halo centered on Virgo~A we want to detect. We decided to apply three different approaches
to remove this curvature from the image: (i) subtraction of the mean radial profile of the map,
(ii) subtraction of a symmetric analytical profile close to the mean radial profile, and (iii)
baselevel correction using an absolutely calibrated map of lower resolution ($35\arcmin$).

\begin{figure}
\resizebox{\hsize}{!}{\includegraphics{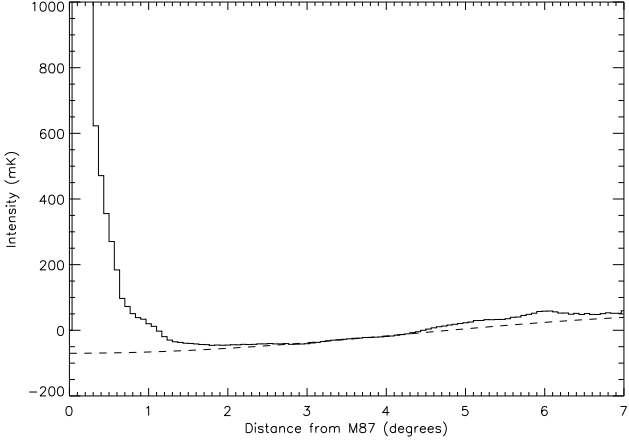}}
 \caption{Radial profiles of the Effelsberg 1.4~GHz map where the point sources are subtracted.
Solid line: mean radial profile. Dashed line: fitted analytical profile (Eq.~\ref{eq:profile}).}
   \label{fig:profiles}
\end{figure}

\begin{figure}
\resizebox{\hsize}{!}{\includegraphics{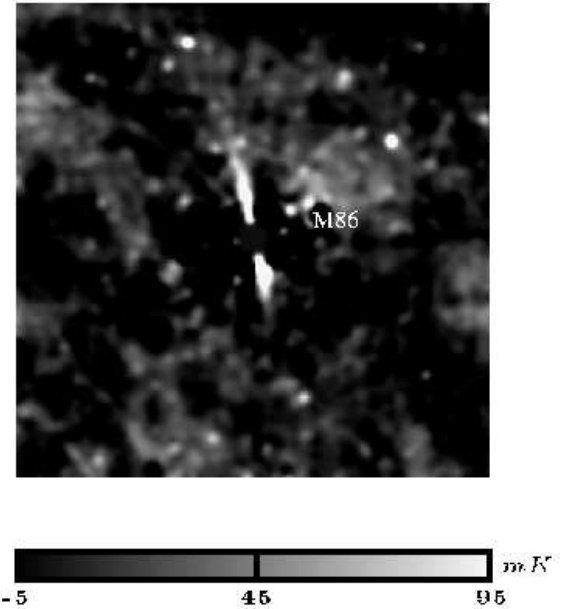}}
 \caption{Effelsberg 1.4~GHz total power map (Fig.~\ref{fig:imap}) where the point sources and the
mean radial profile are subtracted. The resolution is $12\arcmin$. The RA and DEC axis are the same
as in Fig.~\ref{fig:imap}. The location of M86 is indicated.}
   \label{fig:subobservedprofile}
\end{figure}

\begin{figure}
\resizebox{\hsize}{!}{\includegraphics{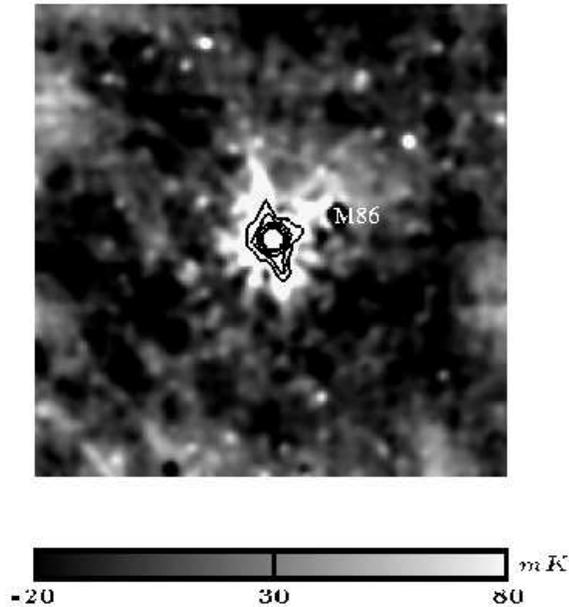}}
 \caption{Effelsberg 1.4~GHz total power map (Fig.~\ref{fig:imap}) where the point sources and the
analytically fitted profile are subtracted. The resolution is $12\arcmin$. The RA and DEC axis are the same
as in Fig.~\ref{fig:imap}. The contours show the radio emission of M87. The location of M86 is indicated.}
   \label{fig:subanalyticalprofile}
\end{figure}

(i)/(ii) We calculated the mean radial profile of the observed image and then fitted an analytical expression
to it
\begin{equation}
S=60\,(1 - \cos(20.8\,D))-70\ ,
\label{eq:profile}
\end{equation}
where $D$ is the distance from M87 in degrees (Fig.~\ref{fig:profiles}) and $S$ is in mK.
These profiles were subtracted from the total power image where the point sources are subtracted, whereas
the sidelobes are still included (Figs.~\ref{fig:subobservedprofile} and \ref{fig:subanalyticalprofile}).
In this way we only investigate the influence of the baselevel subtraction.
Since in the analytic expression the central peak is not included, M87 appears entirely in
Fig.~\ref{fig:subanalyticalprofile}, whereas only its sidelobes remain in Fig.~\ref{fig:subobservedprofile}.
The distributions of the diffuse, extended emission are very similar for both images. The structures that
are elongated in the north-east towards south-west direction most likely belong to the NPS.
A region of diffuse emission can be seen north-west of
M86 with a linear extent of $\sim 1\fdg5$. This feature does not show a north-west elongation typical for the
NPS and thus most probably originates in the Virgo cluster.

\begin{figure}
\resizebox{\hsize}{!}{\includegraphics{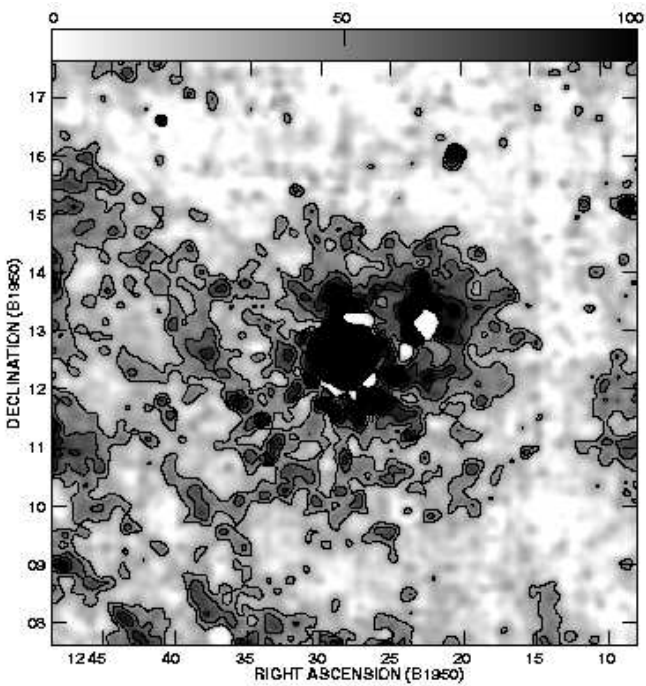}}
 \caption{Effelsberg 1.4~GHz total power map (Fig.~\ref{fig:imap}) where the point sources are subtracted,
the sidelobes are removed, and the large-scale structure is added.
The resolution is $12\arcmin$. The greyscale runs from 0 to 100~mK. The contours are
(1, 1.5, 2, 2.5, 3, 4, 5, 6, 7, 8, 9, 10)\,40~mK. At $12\arcmin$ 1~mK corresponds to 0.77~mJy/beam. }
   \label{fig:finalmap}
\end{figure}

(iii) We use the section of the absolutely calibrated Stockert 1.4~GHz survey map (Reich \& Reich \cite{pr}) to correct
the source-subtracted Effelsberg map first for absolute baselevels and then for flat baselevels across
the $10\degr$ field in order to identify extended emission from the Virgo cluster. The Stockert map
at $35\arcmin$ was first decomposed into its large scale ($\ge 2\degr$) emission using the 'background filtering
method' by Sofue \& Reich (\cite{sofue}). The Effelsberg map was convolved to an angular resolution of $2\degr$
and the difference to the large-scale structures from the Effelsberg map was added to the Effelsberg map at its
original angular resolution. The general gradient in the map from north-west to south-east due to the presence
of the large-scale NPS emission (see Fig.~\ref{fig:imap}) was further reduced by applying again the 'background filtering
method' with a $2\degr$ filter beam to remove the large-scale gradient. Finally small offsets were taken from ring
integration to set the average zero level at about $4\degr$ distance from Virgo~A. The result of this procedure
is shown in Fig.~\ref{fig:finalmap}.
As in Figs.~\ref{fig:subobservedprofile} and \ref{fig:subanalyticalprofile} emission of the NPS has a
characteristic elongation from north-east towards south-west and is visible in the south-eastern quadrant of the image.
At this low surface brightness level the residuals of the sidelobes around M87 can be seen
mainly to the north and south of M87 (compare with Fig.~\ref{fig:subobservedprofile}).
The bright structures surrounding M86 are most probably also due to sidelobes of M86.
However, the large, low surface brightness emission at larger distances from M86 is the same
as the one observed in Figs.~\ref{fig:subobservedprofile} and \ref{fig:subanalyticalprofile}.
It is further discussed in Sect.~\ref{sec:m86halo}.

\subsection{Sidelobe response \label{sec:sidelobes}}

We are interested in weak diffuse emission structures in the surrounding area of Virgo~A and need to know its
sidelobe contribution to a very low level. Deep observations of the antenna characteristics at 1.4~GHz
are difficult
using strong and compact celestial sources like Cas~A, Cygnus~A or Taurus~A, as they have Galactic emission
structures in their surroundings as well as extragalactic
compact sources exceeding the telescope's sidelobe response in intensity and are difficult to separate.
Since no actual antenna pattern is available at 1.4~GHz no full cleaning like that applied by
Rottmann et al. (\cite{rott}) at 10~GHz was possible. Therefore we were limited to remove only the strongest
sidelobe structures from the subreflector support legs as described above, but we will show that the
residual diffraction sidelobes from M87 are weaker than the halo emission we are looking for.

Kalberla et al. (\cite{kalb}) have made antenna pattern measurements of the Effelsberg 100-m telescope at 1.4~GHz
using Cas~A and separated the antenna sidelobes from sky emission by comparing
measurements at different parallactic angles.
The sun was used in addition to measure a small area with the telescope's far--sidelobes at $60\degr$ to
$65\degr$ distance. Kalberla et al.
(\cite{kalb}) measure within $1\degr$ to $4\degr$ distance from the main beam patchy sidelobe structures within $-40$dB
and $-45$dB (their Fig.~2). This level excludes the clearly enhanced cross--like structures from the slit diffraction
pattern of the four feed support legs.

\begin{figure}
\resizebox{\hsize}{!}{\includegraphics{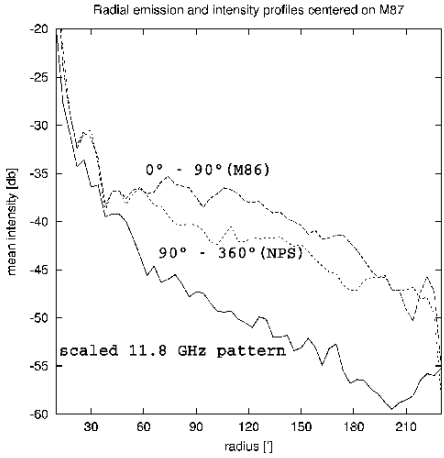}}
 \caption{Radial intensity profiles measured at 11.8~GHz using the EUTELSAT satellite and scaled to 1.4~GHz in
comparison with mean radial intensities measured in the sectors $0\degr - 90\degr$ and $90\degr - 360\degr$
(counterclockwise from north) from the map shown in Fig.~\ref{fig:finalmap}.}
   \label{fig:ring}
\end{figure}

Meanwhile more detailed information on the telescope's sidelobes structures
became available in the course of holographic measurements using geostationary satellites. Their beacons are so strong
that any sky emission does not confuse with the sidelobe structures. These measurements, however, are made at high
frequencies and need appropriate scaling to be comparable with the 1.4~GHz sidelobe structure. We used for comparison
the antenna pattern measured by
Reich \& F\"urst (\cite{rei00}) using the beacon onboard EUTELSAT at 11.7~GHz. The edge taper of the 11.7~GHz feed was lower
than that of the L-band feed to have sufficient sensitivity for the adjustment of the outer telescope panels based on the
holographic result. Therefore the antenna pattern at 11.7~GHz when scaled to 1.4~GHz is expected to show slightly higher
sidelobe levels than present at 1.4~GHz. In addition the telescope surface deviations from the ideal parabolic shape enhance
the high frequency sidelobe levels more than those of the L-band. A comparison of the scaled and ring-averaged 11.8~GHz pattern
with ring-integrated emission from the map shown in Fig.~\ref{fig:finalmap} is displayed in Fig.~\ref{fig:ring}.
We show the emission in
the M86 quadrant (position angles $0\degr$ to $90\degr$).
From the 11.8~GHz pattern the contribution of the four subreflector support legs was
excluded from the ring integration. Clearly the antenna sidelobe levels are below $-40$dB for radii larger than about
$60\arcmin$ and drop to below $-55$dB beyond $180\arcmin$. The emission measured in the M86 area is clearly higher than
any instrumental effect and also higher than the emission measured in the other quadrant, which includes some low-level
filamentary fine structures which are likely from the NPS according to their general orientation. The emission profiles
depend on the adopted zero levels, which we have defined at a radial distance of $40'$ from Virgo~A. The uncertainty
is estimated to about 5~mK, which has an effect of less than 1dB at the $-40$dB level and about 6dB at the $-50$dB
level for example.

\subsection{The M86 radio halo \label{sec:m86halo}}

Based on Figs.~\ref{fig:subobservedprofile}, \ref{fig:subanalyticalprofile}, \ref{fig:finalmap}, and
\ref{fig:ring} the radio halo around M86 is a robust feature of low surface brightness ($\sim 50$~mK or
$\sim 38$~mJy/($12\arcmin$ beam) at a distance of about $1\degr$ from M86).
Its radial extent is $\sim 2\degr$ to the west. It is not possible to determine its extent to the
east because of the emission of M87 and its sidelobe residuals. In Figs.~\ref{fig:subobservedprofile} and
\ref{fig:subanalyticalprofile} the M86 radio halo appears to be asymmetric, which is mainly due to the
way how the radially symmetric emission was subtracted. Thus it is unclear whether the M86 radio halo is
symmetric or not.

For an estimate of its total 1.4~GHz flux we adopted two strategies:
(i) We assume an asymmetric radio halo and integrate the emission in incomplete rings centered on M86 up to a distance of $2\degr$.
The area close to M87 (right ascension offset $< -36'$) was excluded. In this way we derive a lower limit
for the flux density of $4.7 \pm 1.4$~Jy.
(ii) We assume a symmetric Gaussian distribution with a HPW
of $2\degr$ and a peak intensity of 80~mK at $12\arcmin$ resolution.
In this case the halo flux density is about $6.1 \pm 1.2$~Jy for an estimated error of 20\%, which we take
as an upper limit. Combining both results we quote $5 \pm 1.5$~Jy for the halos flux density.

\section{Discussion \label{sec:discussion}}

The analysis of our 1.4~GHz Effelsberg data has shown that the Virgo cluster does not host a bright, large-scale radio halo.
However, we cannot exclude the existence of a weak radio halo centered on M87 of a diameter smaller than
4$\degr$ ($\sim 1.2$~Mpc) with a surface brightness smaller than one third of the radio halo in the Coma cluster.
The detected M86 radio halo is smaller in size and not located in the cluster center as is the
case for the Coma cluster radio halo (see e.g. Deiss et al. 1997). Its radial extent to the west is $\sim 2\degr$.
We cannot determine whether the M86 radio halo is symmetric or not because of its proximity to M87.
It is not excluded that the M86 halo comprises M87 and that the center is located between M86 and M87.
This would resemble the large-scale structure of the X-ray structure (Finoguenov et al. 2004).
As already mentioned in Sect.~\ref{sec:intro} the Virgo cluster is classified as a cooling flow cluster.
Since radio halos avoid this kind of galaxy clusters, this is an expected result.

However, we detect a
radio halo around the elliptical galaxy M86. Despite the complex data reduction (sidelobes, ground emission,
point sources) this result is robust (see Sect.~\ref{sec:data}) and entirely unexpected.
The radio halo appears to be more extended to the north-west (Figs.~\ref{fig:subobservedprofile}, \ref{fig:subanalyticalprofile},
\ref{fig:finalmap}), but the data do not allow to derive a firm conclusion.
Why should M86 harbour a radio halo? If we assume that a radio halo witnesses a past interaction, one possibility
is that the ICM of M86 has interacted with the ICM of M87, i.e. of the Virgo cluster, in the past.
In this case M86 is falling into the Virgo cluster from behind
(its radial velocity is $-1350$~km\,s$^{-1}$ with respect to the cluster mean velocity).

Finoguenov et al. (2004) observed M86 in X-rays with the XMM-Newton telescope. They found a shock
in the south-west of the galaxy center. Since Neilsen \& Tsvetanov (2000), using the method of surface brightness fluctuations,
place M86 at $2.4 \pm 1.4$~Mpc behind the cluster, Finoguenov et al. (2004) concluded that M86 is not interacting with the
Virgo ICM, but with a filament detected in the ROSAT observations (B\"{o}hringer et al. 1994).
Moreover, Finoguenov et al. (2004) claimed that the asymmetric, high surface brightness X-ray distribution of M86 is due to close,
high velocity galaxy--galaxy interactions. They also noted that the X-ray halo around M86 is relatively symmetric,
but there is an offset between the center of the large-scale distribution and the location of M86, which is
displaced to the south-east (their Fig.~12). This offset is consistent with the M86 halo asymmetry we observe in our radio data.
The fact that M87 does not show a bright large-scale halo might indeed indicate that M86 has not
yet passed M87, or that it has passed M87 very recently so that the halo is not built yet.

On the other hand, the distribution of dwarf galaxies (Schindler et al. 1999) around
M87 is asymmetric with its maximum located between M87 and M86. A possible explanation is that M86 has already
passed M87 and that M87 is moving (see Vollmer et al. 2001) or that the dwarf galaxies follow the perturbed gravitational
potential with a certain delay.

We are thus left with two possible scenarios:

(i) M86 is entering the cluster from behind and is located at a distance $D > 2$~Mpc.

(ii) M86 has just passed M87 and a radio halo has not yet been built around M87.

It is not clear how galaxy--galaxy interactions that act on scales between 10~kpc and 50~kpc at most can give
rise to a radio halo of $600$~kpc. In contrast, a large-scale shock in the ICM due to an ICM--ICM
interaction between M86 and M87 can do it. Thus, the radio halo of M86 is most probably due to
an interaction between the ICM of M86 and a large-scale gas structure with a density much smaller than
$10^{-4}$~cm$^{-3}$. The M86 radio halo has a linear extent of
$\sim 0.6$~Mpc. After an interaction, the reverse shock wave that travels with a sound speed
of $\sim 500$~km\,s$^{-1}$ needs $\sim 1$~Gyr to cross the halo. This is about the timescale to
build a radio halo. The radio halo is thus reminiscent of a past interaction of the M86 ICM
with a low density gas, whereas the X-ray morphology is due to a recent interaction of the M86 ICM
with a relatively high density gas.

If M86 was about 2~Mpc away from the cluster center, it would have been about 3~Mpc away
at the time of the beginning of the interaction (assuming a velocity of 1300~km\,s$^{-1}$).
At a distance of 2~Mpc the Virgo ICM density is far below $10^{-4}$~cm$^{-3}$ (Schindler et al. 1999).
Why should there be a gas filament of a density of several $10^{-4}$~cm$^{-3}$?
We do not see how there can be such a large density contrast of the gas that interacts with the M86 ICM
between 3~Mpc (beginning of the large-scale interaction) and 2~Mpc (small-scale interaction).
This density gradient is naturally found further inside the cluster.
Therefore, we prefer scenario (ii), where the interaction began at a distance of $\sim 1$~Mpc to
$2$~Mpc from the cluster center.
This interaction decelerated the outer ICM of M86 whereas M86 itself continued its way into the
Virgo cluster where it now meets Virgo ICM gas densities of several $10^{-4}$~cm$^{-3}$.
It might be near M87, making it move or trailing the dwarf galaxies behind it.
This scenario explains naturally the high ICM density needed to produce the observed shock
and the high velocity galaxy--galaxy encounters needed to explain the high surface brightness
X-ray morphology of M86, since high-speed encounters are more probable in the cluster core
where the galaxy density is higher.

\section{Conclusions \label{sec:conclusions}}

We present Effelsberg 1.4~GHz observations of the central $10\degr \times 10\degr$ of the Virgo cluster, in order
to investigate if the cluster hosts a diffuse radio halo. Three major obstacles have to be overcome
to detect a large-scale, low surface brightness structure: (i) the North Polar Spur (NPS) extends
into the Virgo cluster region, (ii) the high flux density of the central cD galaxy M87 leads to strong sidelobes, and
(iii) the ground radiation due to the large scans. The NVSS survey is used to subtract the point sources
from our total intensity map. In this way we decrease the rms noise level from 7~mJy/beam at a resolution of
$9\farcm4$ to $3$~mJy/beam at a resolution of $12\arcmin$. In the total intensity and polarized intensity
map the NPS appears as elongated structures running along the north-east towards south-west direction
in the south-eastern part of the maps.
Parts of the NPS extend up to the location of M87. We corrected for the non--linear ground radiation in
three ways: (i) we subtracted the observed mean radial profile, (ii) we subtracted an analytical profile,
and (iii) we corrected for the large-scale structure with the help of absolutely calibrated 1.4~GHz observations
of lower resolution. The remaining sidelobe confusion is analyzed using a holographic, deep antenna pattern at 11.7~GHz
using the EUTELSAT satellite.

\noindent
It is concluded that
\begin{enumerate}
\item
we do not detect a bright, large-scale radio halo, as it is observed in the Coma cluster.
\item
We detect a radio halo around the elliptical galaxy M86 with an estimated radial extent of
$\sim 2\degr$ and an estimated total flux density of $5\pm1.5$~Jy.
\item
The detection of the M86 radio halo is robust.
\item
The radio halo is a witness of a past interaction of the M86 ICM with low density gas at the outskirts
of the Virgo cluster.
\item
This leads to a scenario where the ICM of M86 had an interaction $\sim 1$~Gyr ago with a low
density gas, whereas M86 itself undergoes an interaction with a relatively high density gas today
(Finoguenov et al. 2004).
\end{enumerate}

\begin{acknowledgements}
We would like to thank A.~Finoguenov for useful discussions.
\end{acknowledgements}


\begin{thebibliography}{}

  \bibitem[1979]{and} Andernach H., Baker J.~R., von Kap-herr A., \& Wielebinski R. 1979, A\&A, 74, 93

  \bibitem[1971]{elly} Berkhuijsen E.~M., Haslam C.~G.~T., \& Salter C.~J. 1971, A\&A 14, 252

  \bibitem[2002]{boe1} B\"{o}hringer H., \& Sch\"{u}cker P. 2002, in: Merging Processes in Galaxy Clusters,
Ed. L. Feretti, I.M. Gioia, \& G. Giovannini, Astrophysics and Space Science Library, Vol. 272
(Dordrecht: Kluwer) p. 133

   \bibitem[1994]{boe2} B\"{o}hringer H., Briel U.~G., Schwarz R.~A. et al. 1994, Nature, 368, 828

   \bibitem[1976]{brouw} Brouw W.~N., \& Spoelstra, T.~A.~Th. 1976, A\&AS, 26, 129

   \bibitem[1998]{con} Condon J.~J., Cotton W.~D., Greisen E., et al. 1998, AJ, 115, 1693

   \bibitem[1997]{dei} Deiss B.~M., Reich W., Lesch H., \& Wielebinski R. 1997, A\&A, 321, 55

   \bibitem[1996]{ebe} Ebeling H., Voges W., B\"{o}hringer H., et al. 1996, MNRAS, 281, 799

   \bibitem[1988]{emerson88} Emerson D.~T., \& Gr\"ave R. 1988, A\&A 190, 353

   \bibitem[2004]{fino} Finoguenov A., Pietsch W., Aschenbach B., \& Miniati F. 2004, A\&A, 415, 415

   \bibitem[2001]{giovf} Giovannini G., \& Feretti L. 2002a, in: Merging Processes in Galaxy Clusters,
Ed. L. Feretti, I.~M. Gioia, \& G. Giovannini, Astrophysics and Space Science Library, Vol. 272
(Dordrecht: Kluwer) p. 197

   \bibitem[2001]{giov} Giovannini G., \& Feretti L. 2002b, in: Highlights of Astronomy, Vol. 12,
as presented at the XXIVth General Assembly of the IAU - 2000 [Manchester, UK, 7 - 18 August 2000].
Ed. H. Rickman (San Francisco: Astronomical Society of the Pacific) p. 513

   \bibitem[1980]{kalb} Kalberla P.~M.~W., Mebold U., \& Reich, W. 1980, A\&A, 82, 275

   \bibitem[2000]{neil} Neilson E.~H., \& Tsvetanov, Z.~I. 2000, ApJ, 536, 255

   \bibitem[2000]{owen} Owen F.~N., Eilek J.~A., \& Kassim N.~E. 2000, ApJ, 543, 611

   \bibitem[1986]{pr} Reich P., \& Reich W. 1986, A\&AS, 63, 205

   \bibitem[1997]{pr97} Reich P., Reich W. \& F\"urst E. 1997, A\&AS, 126, 413

   \bibitem[2000]{rei00} Reich W., \& F\"urst E. 2000, Kleinheubacher Berichte, 43, 175

   \bibitem[1997]{roet} R\"{o}ttgering H.~J.~A., Wieringa M.~H., Hunstead R.~W., \& Ekers R.~D.
1997, MNRAS, 290, 577

   \bibitem[1996]{rott} Rottmann H., Mack K.-H., Klein U., \& Wielebinski R. 1996, A\&A, 309, L19

   \bibitem[1999]{schin} Schindler S., Bingelli B., \& B\"{o}hringer H.  1999, A\&A, 343, 420

   \bibitem[1979]{sofue} Sofue Y., \& Reich W. 1979, A\&AS, 38, 251

   \bibitem[1972a]{titusa} Spoelstra T.~A.~Th. 1972a, A\&AS, 5, 205

   \bibitem[1972a]{titusb} Spoelstra T.~A.~Th. 1972b, A\&A, 21, 61

   \bibitem[1984]{tul} Tully R.~B. \& Shaya E.~J. 1984, ApJ, 281, 31

   \bibitem[1998]{bu98} Uyan{\i}ker B., F\"urst E., Reich W., Reich P., \& Wielebinski R. 1998, A\&AS, 132, 401

   \bibitem[1999]{bu99} Uyan{\i}ker B., F\"urst E., Reich W., Reich P., \& Wielebinski R. 1999, A\&AS, 138, 31

   \bibitem[1993]{w93} White S.~D.~M., Briel U.~G., \& Henry J.~P. 1993, MNRAS, 261, L8

   \bibitem[1988]{whi} White R.~E. \& Sarazin C.~L. 1988, ApJ, 335, 688

\end{thebibliography}
\end{document}